\documentclass[journal]{vgtc}                
\ifpdf
  \pdfoutput=1\relax                   
  \usepackage{graphicx}                
  \DeclareGraphicsExtensions{.pdf,.png,.jpg,.jpeg} 
\else
  \usepackage{graphicx}                
  \DeclareGraphicsExtensions{.eps}     
\fi%

\graphicspath{{figures/}{pictures/}{images/}{./}} 

\usepackage{microtype}                 
\PassOptionsToPackage{warn}{textcomp}  
\usepackage{textcomp}                  
\usepackage{mathptmx}                  
\usepackage{times}                     
\usepackage{cite}                      
\usepackage{tabu}                      %
\usepackage{algorithm}
\usepackage{algorithmicx}
\usepackage{algpseudocode}
\usepackage{graphicx}
\usepackage{amsmath}
\usepackage{booktabs}                  

\usepackage{textcomp}
\usepackage{hyperref}
\usepackage{animate}
\usepackage{scalerel}
\usepackage{tikz}
\usepackage{svg}
\usepackage{enumitem}
\usepackage{booktabs}
\usepackage{array}
\usepackage{multirow}

\onlineid{0}

\vgtccategory{Research}
\vgtcpapertype{please specify}

\title{Regshock: Interactive Visual Analytics of Systemic Risk in Financial Networks}

\author{Zhibin Niu, Junqi Wu, Dawei Cheng, and Jiawan Zhang}
\authorfooter{
\item
Zhibin Niu, Junqi Wu, and Jiawan Zhang are with College of Intelligence and Computing,Tianjin University. E-mail: zniu@tju.edu.cn.

}

\shortauthortitle{Biv \MakeLowercase{\textit{et al.}}: Global Illumination for Fun and Profit}

\abstract{Financial regulatory agencies are struggling to manage the systemic risks attributed to negative economic shocks. Preventive interventions are prominent to eliminate the risks and help to build a more resilient financial system. Although tremendous efforts have been made to measure multi-risk severity levels, understand the contagion behaviors and other risk management problems, there still lacks a theoretical framework revealing what and how regulatory intervention measurements can mitigate systemic risk. Here we demonstrate  \texttt{regshock}, a practical visual analytical approach to support the exploration and evaluation of financial regulation measurements. We propose \texttt{risk-island}, an unprecedented risk-centered visualization algorithm to help uncover the risk patterns while preserving the topology of financial networks. We further propose \texttt{regshock}, a novel visual exploration and assessment approach based on the simulation-intervention-evaluation analysis loop, to provide a heuristic surgical intervention capability for systemic risk mitigation. We evaluate our approach through extensive case studies and expert reviews. To our knowledge, this is the first practical systemic method for the financial network intervention and risk mitigation problem; our validated approach potentially improves the risk management and control capabilities of financial experts.
} 


\CCScatlist{ 
 \CCScat{K.6.1}{Management of Computing and Information Systems}%
{Project and People Management}{Life Cycle};
 \CCScat{K.7.m}{The Computing Profession}{Miscellaneous}{Ethics}
}

\teaser{
  \centering
  \includegraphics[width=\linewidth]{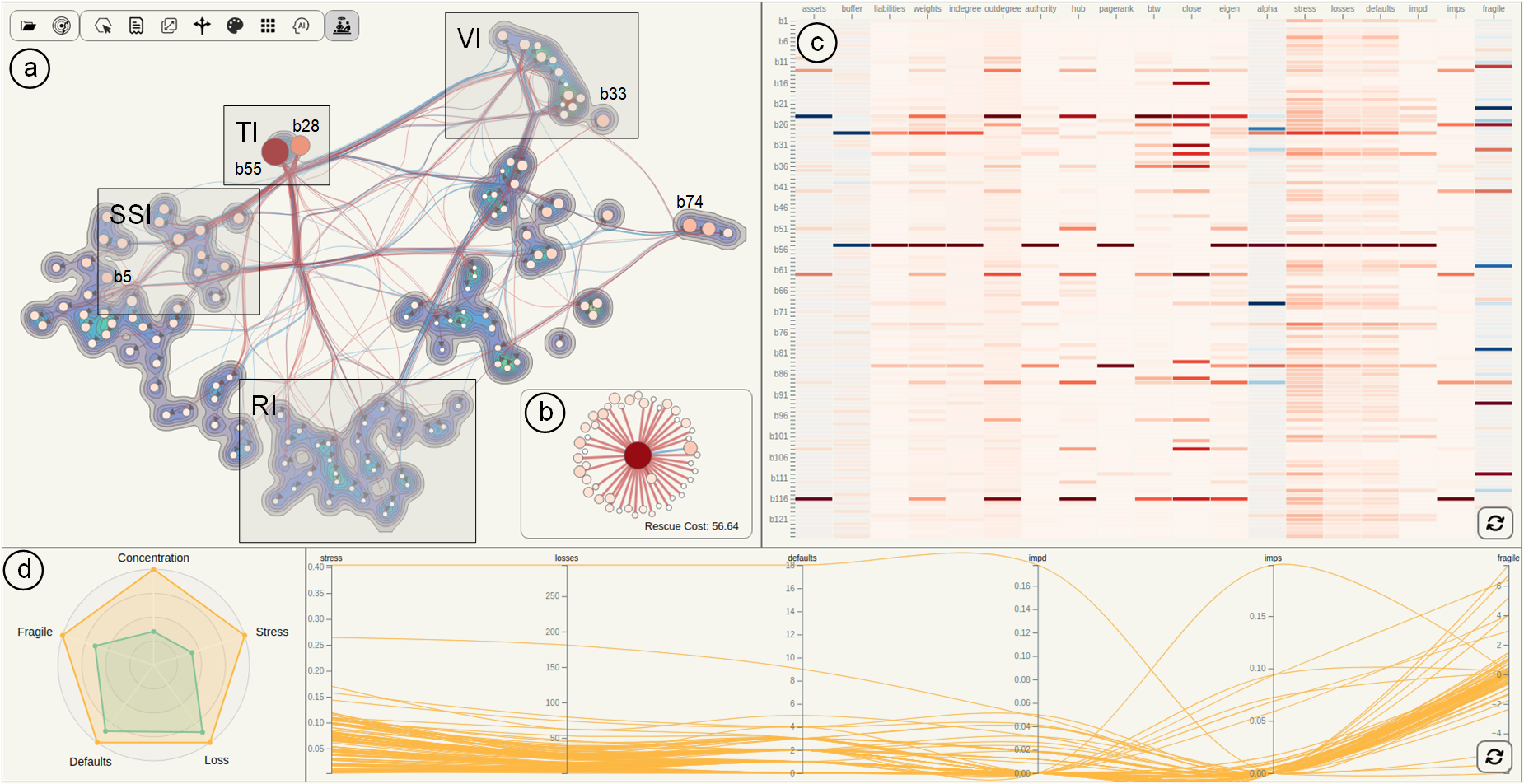}\vspace{-8pt}
  \caption{The system interface of \texttt{regshock}: a) Risk-island view, uncovering multi-risk patterns of the financial network, providing risk exploration and mitigation intervention tools; b) Operating table view, providing risk mitigation playground. c) Detail View, visualizing the finest-grain information of the financial entities. d) Intervention assessment view, visualizing risk mitigation effects to identify optimal operation strategy.}
	\label{fig:teaser}
}

\vgtcinsertpkg

\begin{document}


\firstsection{Introduction}

\maketitle

Effective regulation of financial networks received considerable attention after the global financial crisis~\cite{cottle2008global}. The greatest threat to systemic risk is the unexpected shocks and their contagion behavior in financial networks~\cite{bardoscia2017pathways}. In case of a sudden economic shock, defaults can spread like wildfire and be amplified in both reach and impact, leading to large-scale failures~\cite{catanzaro2013network}. Effective regulations, especially preventive measures, are critical to avoid enormous damage. The stakeholder (financial regulatory agencies and financial entities) urge the adoption of advantage regulation technology (Regtech) to manage the economy better and avert future disaster by reducing the potential impact of ``black swan events''~\cite{allen2000financial,glasserman2015likely,gai2010contagion,poledna2016elimination}.

Regulating the systemic risks requires an interdisciplinary effort of financial and computing. Tremendous efforts are made to regulations such as enhance the transparency and compliance audit, measure multi-aspect risk severity, identify systemically important nodes~\cite{del2020multiplex}, understand the contagion behaviors such as susceptibility and vulnerability~\cite{chiu2015industry,peckham2013economies,cifuentes2005liquidity}, estimate the resilience of these systems to events such as financial shocks, crises, and cascade failures~\cite{anand2013network, baranova2017simulating,cimini2015systemic}, and many others. However, the mechanisms that might cause systemic risk are still unclear due to the nonlinear aspects of financial contagion~\cite{catanzaro2013network}. Moreover, these economic models usually focus on specific ad-hoc problems, thus infeasible to provide a complete theoretical framework of how regulatory interventions can mitigate risks~\cite{catanzaro2013network}. Visual analytics empowers experts' sense-making capability for better decision-making through human-machine co-operation, which naturally matches the complex financial analysis demands~\cite{haldane2014managing, niu2020iconviz, niu2020regvis}. A growing number of financial visualization research are published these years~\cite{haldane2013rethinking,chang2007wirevis,yue2018bitextract,leite2017eva,ko2016survey,xie2014vaet}. However, most of them concentrated on utilizing visual analytics to enhance the economic systems' understanding through exploration analysis~\cite{Heijmans2014Dynamic,ronnqvist2015bank,niu2018visual}, few have considered practical decision-making for the risk mitigation problem.

Traditional network visualization technologies are insufficient to facilitate the advanced risk analysis of financial networks owing to their unpredictable, dynamic, and complex characters. The financial system are usually modeled as temporal networks composed of high-dimensional attributed entities and their sophisticated links. Traditional ones such as force-directed graph layout are good at presenting the information. Still, they are insufficient to assist complex business analysis such as multi-facet risk analysis of the temporal network, contagion effects assessment, and prescriptive decisions such as proactive interventions for risk mitigation. The complexity of practical business analysis requires us to emphasize more from the analysis task aspect.


We have identified that practical intervention measures for risk mitigation can be fulfilled through the heuristic approach of firstly exploration and then validation by co-work with financial experts. Based on the hypothesis, we propose a practical visual analytical methodology \texttt{regshock} of \emph{simulation}-\emph{intervention}-\emph{evaluation} to support the exploration and evaluation of regulation measurements. To our knowledge, this is the first practical systemic method for the financial network intervention and risk mitigation problem. The main contributions of this work are as follows:

\setlist{nolistsep}
\begin{itemize}  [noitemsep, leftmargin=*]
    \item  We propose an unprecedented risk-centered network layout algorithm and \texttt {risk-island} visualization to facilitate uncovering the risk patterns while preserving topology in financial networks. The \texttt {risk-island} visualization design can better support complex financial network analysis than traditional ones.

    \item  We propose a novel visual analytic approach \texttt {regshock} to facilitate the heuristic surgical-intervention capability for risk mitigation in financial networks. It provides sufficient flexibility, intuition, and interpretability through simulation-intervention-evaluation analysis loops for practical risk mitigation.
\end{itemize}





\section{Related Work}
\label{relatedwork}
We begin by briefing the background, then survey the progress of visualization and visual analytics in financial regulation, and finalize it by summarizing our research's uniqueness.

\subsection{Financial network}
The two most typical financial networks are interbank lending market~\cite{angelini2011interbank,afonso2011stressed}  and networked-loans~\cite{niu2018visual,niu2020iconviz,wang2020evolution}. The interbank lending market is a market in which banks lend money to each other for a specified term. Most interbank loans have a maturity of one week or less, and most of them have a maturity of over one day. Such loans are made at the interbank rate (also called the overnight rate if the loan term is overnight). A sharp decline in the volume of transactions in this market was a major factor in the collapse of several financial institutions during the 2007-2008 financial crisis.
The banking industry initiated risk quantification model started from the 1950s~\cite{hodgman1960credit,freimer1965bankers}. Thirty years ago, the Basel Committee issued a series of recommendations on banking regulations (Basel I, II, and III) to enhance the understanding of critical regulatory issues and improving macro-prudential oversight~\cite{benzin2003approaches,eubanks2010status}. The Basel Capital Accord has been widely accepted by banks around the world \cite{montgomery2005effect,benink2002new}. Some indicators (e.g., the probability of default loss-given default, exposure at default, etc.) have been introduced to help manage the risk~\cite{altman2005effects}. However, such methods highly depend on the financial statements of the individual company and may not be well-suited for networked-loans because the network relationship is unique and exceeds the original hypothesis~\cite{meng2017netrating,niu2020iconviz}. The networked-loans were a byproduct of the stimulus program after the global financial crisis when several central banks loosened credit standards to help endangered businesses survive in the crises. Corporations are allowed to serve as guarantors for one another when applying for loans from commercial banks (to enhance security) and thereby forming complex financial networks~\cite{cong2019credit,jiang2015vulgarisation}.

\subsection{Regulatory visualization}
We provided http://regvis.net, a visual bibliography of regulatory visualization for better indexing the literature~\cite{niu2020regvis}. The rapid development of regulatory technology has raised awareness of information visualization and visual analytics in this area. However, so far, only a few visual analytics solutions target the data analysis tasks in financial scenarios owing to the financial domain complexity. The financial community primarily utilizes the graphics such as bar charts, Box plots, map charts, and other basic ones to present discoveries~\cite{vlab}. There is some valuable attempt in the interdisciplinary research of computing and finance. For example, the self-organizing map was introduced to encode the financial stabilities~\cite{visklab,sarlin2016macroprudential,sarlin2011visual,sarlin2013mapping,sarlin2013self}. Analytics method deeply tied to the business such as stock analysis and fraud detection attracted relative intensive attention~\cite{dumas2014financevis}. Classic visual analytics systems for such applications, including monitoring stock trades and quotations~\cite{kirkland1999nasd,leite2018eva}, detect suspicious accounts, transactions, behaviors through wirevis approach~\cite{chang2007wirevis}, identify unusual trading patterns, suspected traders (i.e., attackers), and attack plans through 3D tree maps~\cite{huang2009visualization}.

Network visualization is employed to represent bank interrelations through financial discussion data~\cite{constantin2018network}. The force-directed layout is perhaps the most extensively utilized in the financial area. Moreover, the network centrality measurements such as node degree, betweenness, and closeness, K-core shell measure and visualize the node or edge importance. For example, Rönnqvist and Sarlin collected text data from online financial forums and generated and visualized a co-mentioned bank network (i.e., interbank network), with which to quantify the bank interdependence (using centrality measurements), such as interbank lending and co-movement in market data~\cite{ronnqvist2015bank}. Financial criminal networks can be partitioned into subgroups of individuals by the centralities in their network~\cite{didimo2011an}. Bottom-up and top-down interaction are demonstrated can be effective in revealing financial crimes such as money laundering and fraud in the financial activity network~\cite{didimo2011advanced}. Heijmans and others used animation to visualize and analyze the large transaction networks in the daily Dutch overnight money market~\cite{Heijmans2014Dynamic}. There were some other publications mining the subgraph structures and patterns to interpret the financial meaning. Among them, BitExTract was developed to observe the evolution of transaction and connection patterns of Bitcoin exchanges from different perspectives~\cite{yue2018bitextract}. An ego-centered node-link view depicts the trading network of exchanges and their temporal transaction distribution and facilitates the recognition of unique patterns. We have contributed to the first visual analytics method for the networked-loans problem~\cite{niu2018visual}, followed by a serial of data-driven approach for finding systemically important institutions~\cite{cheng2019dynamic,cheng2019risk} and understanding contagion behaviors~\cite{cheng2020contagious, niu2020iconviz}.

This work distinguishes itself significantly from all these previous work by step forward to facilitate the preventive response's prescription capability. As aforementioned, there lacked a theoretical method that reveals what and how regulatory intervention can disrupt shock propagation, thus infeasible to facilitate practical preventive measures. The interactive visual methodology can pave a pathway for better management of the economy and a means to avert future disaster.

\section{Problem Statement and Business Modeling }
We worked closely with financial risk management experts to better understand the real-world financial challenges. 
Our ultimate goal is to provide practical intervention capabilities for the financial system risk mitigation. Effective intervention requires in-depth understanding the multi-risks in the financial networks especially the effect of unexpected negative shocks, assessment and comparison of different interventions. To facilitate optimal intervention analysis, we sum up the risk mitigation requirements from the business aspect, describe the multi-risk modeling for financial networks, and the shock and intervention effect modeling.

\subsection{Risk Mitigation Requirements} 
\label{riskquetions}

The financial network is a special kind of complex network generated when several financial entities such as traders, firms, banks, and financial exchanges are linked together through transactions, guarantee, stock or bond ownership~\cite{gale2007financial,niu2018visual}. In terms of network science, financial networks consist of financial nodes representing financial institutions or participants and edges representing formal or informal relationships between nodes~\cite{allen2009networks}. The defaults can be amplified through the financial network. Financial regulatory agencies urge preventive measures to regulate the systemic crisis risk response to economic shocks. Regulating the financial networks requires systematic treatment strategy. We summarize risk mitigation requirements as follows:

\textbf{Multi-risk assessment.} There are multiple source of risk (from individual level to systemic level) hidden in the financial networks. The systemic risk is the consequences of economic shocks~\cite{schwarcz2008systemic,anabtawi2011regulating}. Unlike the epidemic spreading in which the virus usually spreads to only adjacent nodes, the default or stress of financial institutions attacked by economic shock will immediately trigger large-scale default or stress of other entities, leading to massive system failures. Particular attention should be paid to anticipate how debt default may spread for appropriate response interventions to prevent large-scale defaults. Accidental default is usually tolerable, while large-scale defaults or systemic financial crises must be prevented.

\textbf{Preventive measures.} Although tremendous efforts are made to measure multi-aspect risk severity, understand the contagion behaviors such as susceptibility and vulnerability, and others~\cite{bardoscia2015debtrank, silva2017monitoring, cheng2020delinquent}, there still lacks a theoretical framework suggesting how and what regulatory intervention can disrupt shock propagation, thus infeasible to facilitate practical preventive measurements. Intervention measurements should be taken case by case owing to the complexity of financial network problems. Thus, comparative intervention assessments are critical for optimal risk treatment measures.


\subsection{Risk Modeling}
\label{secfeatures}
We consider the risk-centered graph  as $G(f)=(V,E)$, Let $G$ have $n$ nodes, denoted by $V={v_{1}, v_{2},...,v_{n}}$, where $v_{1}$ is the source node and all the other nodes are those linked from $v_{1}$ through debt relationships. The structure of $G$ is defined by its adjacency matrix $A$ where $a_{ij}=1$ indicates an nontrivial fluidity link. The objective is arrange a layout following above criteria, computing node placement and edge layout.

We employ quantified multi-risk metrics ($FR=[r_{b},r_{n}, r_{i}, r{s}]$). As detailed in Table~\ref{features}, they fall into the following four categories:
\setlist{nolistsep}
\begin{itemize} [noitemsep, leftmargin=*]
\item \textbf{Balance sheet.} The balance related financial profiles relate closely to the risk levels, as the Basel Capital Accord (see Section~\ref{relatedwork}) indicated.
\item \textbf{Traditional network centralities.} Traditional centrality measures such as degrees, betweenness, closeness, eigen, and alpha centrality from network theory is been proven effective for measuring the importance of nodes in financial networks~\cite{niu2018visual}.
\item \textbf{Entity risk indicators.} All financial entities have systemic importance owning to the mutually beneficial network structure. We employ three metrics (fragility, impact diffusion, and impact susceptibility) to measure the of severity of impacts to the system.
\item\textbf{Systemic financial indicators.} They measures the shock impact to the entire financial system.
\end{itemize}

\begin{table}[!tbh]
\center
\begin{tabular}{@{}ll@{}}
\toprule
Category                & Risk indicator                                                                                                                                                         \\ \midrule
Balance sheet ($r_{b}$)            & Assets*                                                                                                                                                                \\
                        & Liabilities                                                                                                                                                            \\
                        & Capital Burffer*                                                                                                                                                       \\
                        & Weight                                                                                                                                                                 \\
Traditional centralities ($r_{n}$)   & Degree (in/out)                                                                                                                                                        \\
                        & Authority and Hub, Pagerank                                                                                                                                            \\
                        & K-shell                                                                                                                                                                \\
                        & \begin{tabular}[c]{@{}l@{}}betweenness, closeness,\\ Eigen centrality, \\ Alpha centrality\end{tabular}                                                                \\
Entity risk indictors ($r_{e}$)         & Fragility~\cite{battiston2012debtrank}*                                                                                                                                \\
                        & Impact diffusion~\cite{silva2017monitoring}*                                                                                                                           \\
                        & Impact susceptibility~\cite{silva2017monitoring}*                                                                                                                      \\
Systemic risk indictors ($r_{s}$) & System fragility~\cite{battiston2012debtrank}*                                                                                                                         \\
                        & \begin{tabular}[c]{@{}l@{}}System stress, System loss, \\ and System default ~\cite{bardoscia2015debtrank,battiston2015leveraging,battiston2013systemic}*\end{tabular} \\
                        & System concentration~\cite{battiston2012debtrank}*                                                                                                                     \\ \bottomrule
\end{tabular}
\vspace{5pt}
\caption{The multi-risk factors for modeling the financial risk in the interbank network (see definitions in appendix). The factors with * could be altered by external shocks, refer to Figure~\ref{updating} for more details.}\vspace{-20pt}
\label{features}
\end{table}

\subsection{Shock and Intervention Modeling}
\label{shocksection}
\begin{figure}[!tbh]
  \centering
  \includegraphics[width=1\linewidth]{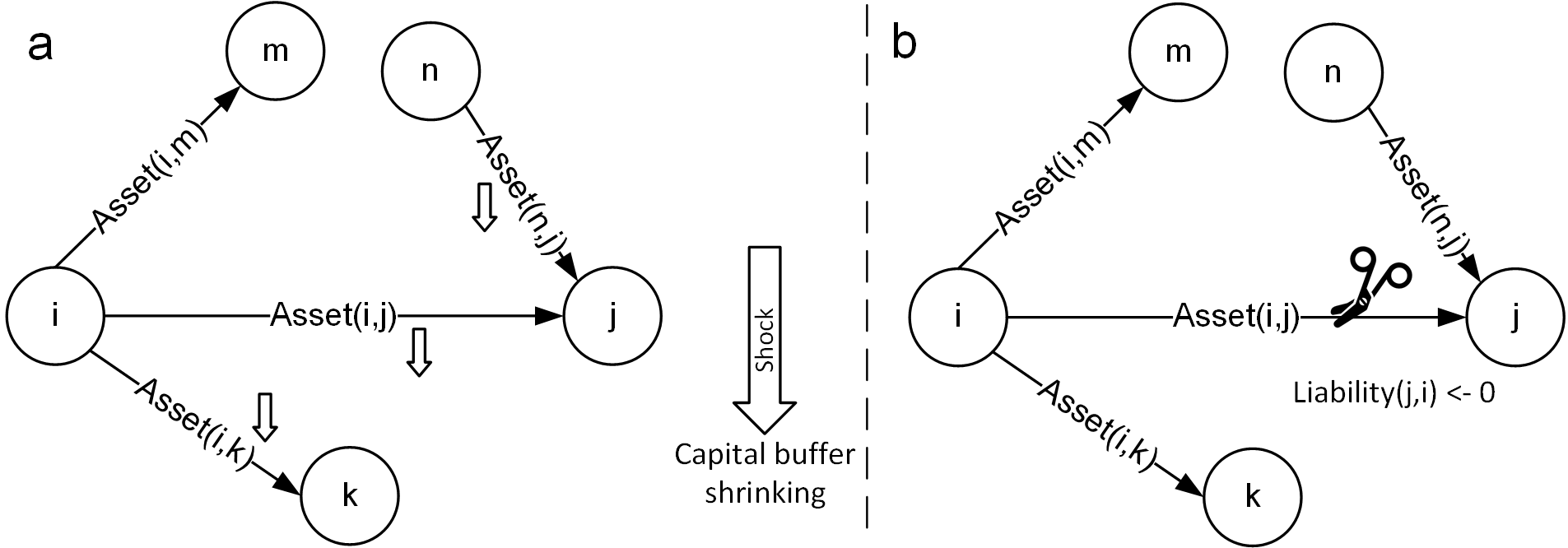}\vspace{-10pt}
  \caption{Shock and Intervention modeling schematic}\label{updating}\vspace{-12pt}
\end{figure}

The shock-based pressure test is a necessary component to assess a financial network's ability to withstand risks. We employ three mainstream shock-response models in the banking industry to simulate the contagion process and estimate the system's default propagation effect and impact. The contagion algorithms we use include: 1) Threshold Propagation (default cascades); 2) Linear Propagation~\cite{bardoscia2015debtrank,battiston2012debtrank}; 3) Combination of threshold and linear propagation. Figure~\ref{updating}(a) gives an illustrative case on how we update the financial and risk attributes over the shock. When the interbank network suffer shocks, the banks can use their capital as a buffer to absorb the shocks. When a bank suffers a shock, its capital buffer will first shrink at a certain proportion, then the shock will propagates some losses of its debts to its creditors; and successively spread across the network. We simplify this process to the procedure as shock to buffer to assets.

The Intervention to the financial network include highly flexible tasks including remove node of interest, partition the network, replace the node with specified one, and many others. In this paper, we simulate the partition effect with a similar method to~\cite{afonso2011stressed}. As Figure~\ref{updating}b shows, when a specific node was removed from the interbank network, the liability of the removed node will be cleared. And its creditors will loss corresponding assets. We simplify this process to the procedure as partition to assets. It is noted that we only give illustrative case here, more details please refer to the literature~\cite{bardoscia2015debtrank,silva2017monitoring,silva2016network, anand2015filling,cinellinetworkriskmeasures}

\begin{figure*}[!tbh]
  \centering
  \includegraphics[width=1\linewidth]{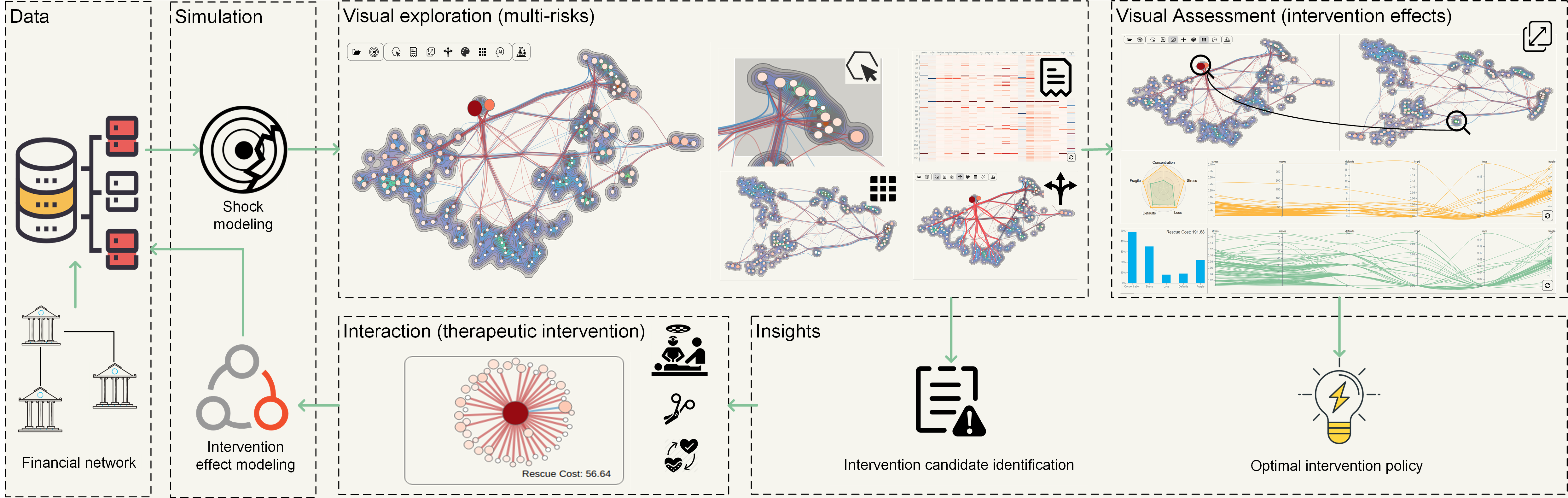}\vspace{-8pt}
  \caption{Method of \texttt{regshock}. It includes visual exploration of multi-risks, intervention interactions, shock and intervention effect simulation, and visual assessment for risk mitigation.}\label{method}\vspace{-10pt}

\end{figure*}
\section{Method}
This section presents our visual analytics method to facilitate the risk mitigation analysis. To meet the business requirements, we first frame the design requirements to support risk mitigation analysis capability, and then derive the design decisions to better fulfill them. After that, we overview our \texttt{regshock} approach that addresses the requirements and design decisions. Immediately next to that, we describe the \texttt{risk-island} layout that was designed to meet the sophisticated multi-risk analysis requirements, followed by details of other coordinated functional views.


\subsection{Requirements Analysis}
To facilitate the risk mitigation analysis, the visual analytics should support the following analysis capability:
\setlist{nolistsep}
\begin{description} [noitemsep, leftmargin=*]
\item[R.1] \textbf{Intuitive and interpretable multi-risk visualization and exploration.} The practical financial network involves complex multi-aspect risks (internal or external shock or other events). Intuitive and interpretable multi-risk visualization could enhance the understanding of the risk type and help to identify risk source, and thus facilitate intervention of risk mitigation.
  \item[R.2] \textbf{Flexible shock and intervention effect modeling and simulation.} The stress test is based on the shock simulations. Although the modeling and simulation is not directly visualized, they are the necessary components to support hypothesis validation. Thus therefore, we should have a reasonable and extensible mechanism to configure external shocks.
  \item[R.3] \textbf{Facilitate the users to apply intervention operations to the financial network interactively.} At present it could be difficult for the financial experts to verify their intervention operations. The real-world financial networks could be composed of hundreds or thousands of nodes and the experts usually have no choice to do micro-economic experiments but only focus on macro-economic problems. Interactive intervention capability have the potential to support financial experts to explore their intervention ideas.
  \item[R.4] \textbf{Facilitate intervention effect assessment and strategy comparison to find optimal one.} A practical risk mitigation should include intervention effect assessment and strategy comparison to support financial experts make decisions.
\end{description}

The above analysis capability requires specially tailored visualization and interaction design for the financial networks owing to the complexity of analysis tasks and sophisticated data. In response, we derive a series of design decisions addressing, as outlined below.
\setlist{nolistsep}
\begin{description} [noitemsep, leftmargin=*]
   \item[D.1] \textbf{The map metaphor for financial network visualization and earthquake metaphor to negative economic shock (R.1). } The financial network suffering external shocks are often described as a financial tsunami or a financial earthquake. Analogy, we choose to use the map as the metaphor of financial network and use earthquake as the metaphor of negative economic shock. Besides, we use the island as metaphor of financial entities of similar risk level. 

    \item[D.2]\textbf{Surgery metaphor for intervention operation to the financial network (R.3).} The intervention to the financial network is like the surgery operations, including careful examinations, rehearsal of different scenario, surgery effect evaluation, and others. So we choose to use the surgery metaphor for the design of intervention operations. The surgery could be exploratory surgery, therapeutic surgery, excision, transplant, reconstructive surgery. Similar intervention to the financial network could be effective to the risk mitigation in financial networks .

 \item[D.3]\textbf{Simulation-Intervention-Evaluation analysis methodology (R.2, R.3).} A practical risk mitigation method should facilitate the users to apply intervention operations to the financial network interactively based on risk exploration, simulation, intervention, and evaluation analysis loop.

  \item[D.4] \textbf{A confluence of multi-view interactions for optimal strategy exploration (R.4).} Optimal intervention operations can be achieved through coordinated multi-view interactions to facilitate the simulation-intervention-evaluation capability. Appropriate multi-view designs help experts reduce the visual burden and improve their understanding of the actual situation.
\end{description}

\subsection{Overview of the Approach}
Figure~\ref{method} shows the method of \texttt{regshock}. It mainly include four modules to fulfill the \emph{simulation-intervention-evaluation} analysis loop.

The financial data of related entities are initially collected, reconstructed into financial networks, and stored. To estimate the multi-risk, we apply parameterized-negative shock (refer to Section~\ref{shocksection}) to the entire financial system for the stress test~\footnote{Stress test, financial terminology, is an analysis or simulation designed to determine the ability of a given financial instrument or financial institution to deal with an economic crisis.}. The shock will affect $r_p, r_e, and\ r_s$, and the initial financial network $FN_{o}$ is updated by the backend shock modeling algorithm to $FN_{s}$.

\begin{figure}[hbt!]
  \centering
  \includegraphics[width=1\linewidth]{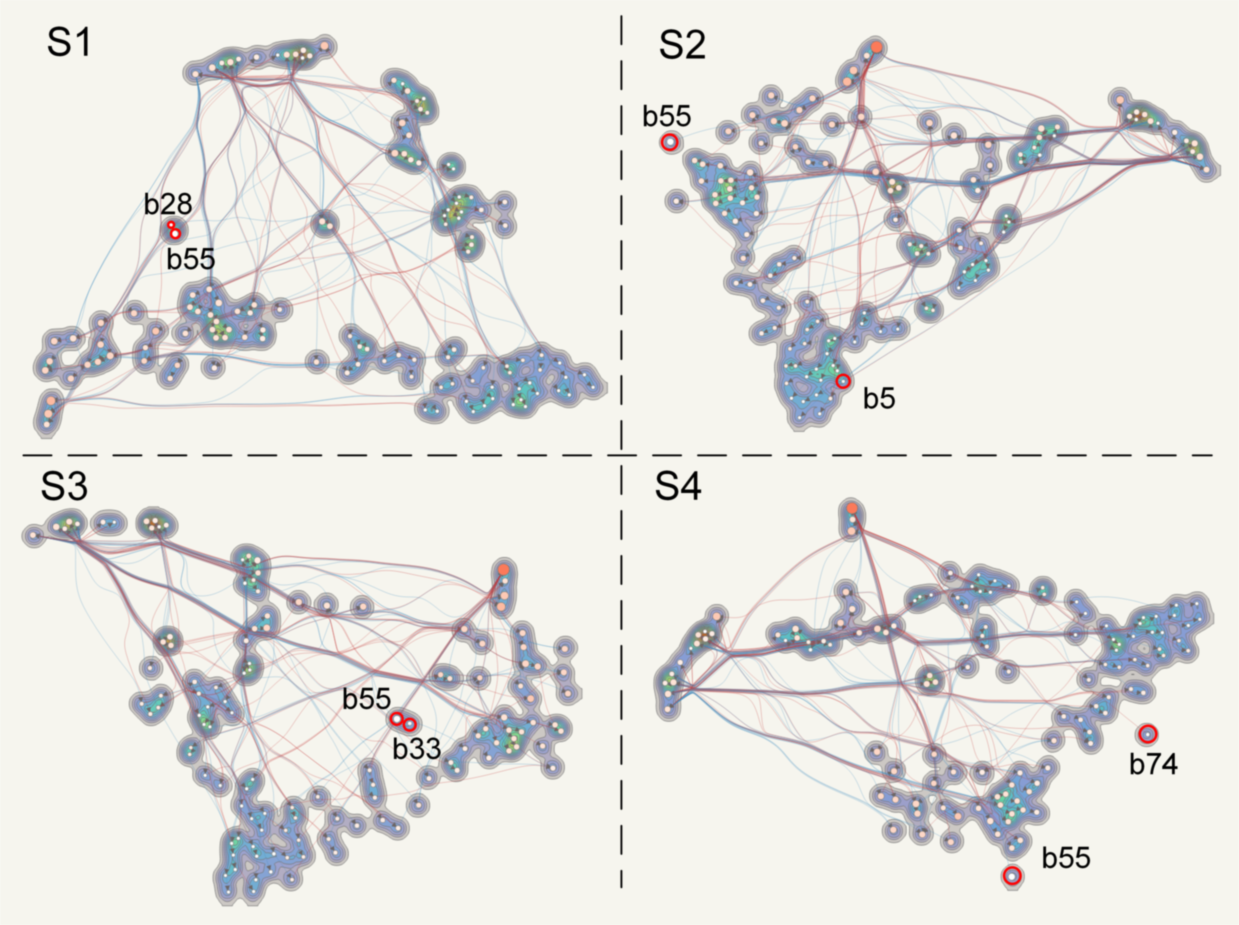}\vspace{-8pt}
  \caption{The risk-islands of the intervention operations S1-S4. The overall node distribution and concentration levels see a significant adjustment.}
  \label{case2}\vspace{-8pt}
\end{figure}

The visual exploration of multi-risks is started from visualizing the $FN_{s}$ network through \texttt{risk-island} that will be detailed in the next section. The risk-island view can seamlessly characterize the risk models while preserving the network topology (R.1). Each island presents financial entities with similar risks; the visual encoding of nodes and edge provides visual hints for risk mitigation; users can conduct partition operations on edges to verify their risk mitigation strategies. Users are able to use the interaction tools (see Section~\ref{tools}) to discovery various risk patterns of risk-islands, view node details or fluidity (see Section~\ref{case1}), or view the intervention result (see Section~\ref{case2}). Through the visual exploration of multi-risks, the financial experts may gain insights into the overall risk levels and have some hypotheses on how to intervene in the network to reduce risks. They may obtain an intervention candidate list as the insight.

With the intervention candidate and intervene hypothesis, the users can conduct ``therapeutic intervention''  on the operating table (Figure~\ref{fig:teaser}b), conducting excision, transplant, and many other operations. The intervention operation will affect $r_p, r_n, r_e, and r_s$, and the backend intervention effect modeling algorithm will update the financial network $FN_{s}$ as $FN_{e}$.

We can apply shock simulation to the financial network $FN_{i}$ again; the generated new financial network is annotated as $FN_{is}$. Now, users are able to assess the intervention effect in the visual assessment view. They can compare the difference of risk-islands, compare the difference of systemic and bank-level risks. It is noted that the analysis is heuristic and iterative. Users are able to verify their intervention hypothesis until they reach an optimal one.

\subsection{Visualization and Interaction Design}
\label{s:interface}
In this section, we describe the visualization and interaction design of the \texttt{regshock} approach.

\subsubsection{Risk-island Visualization}
Facilitating advanced risk analysis of the financial systems faces two-fold challenges. Firstly, the financial systems are unpredictable, dynamic, and complex~\cite{haldane2013rethinking, boot1997financial} and the related risk analysis is also complex and diverse. They could range from entity/systemic multi-risk analysis, early detection and risk assessment of potential system loss, preventive measures. This requires us to facilitate aggregation, multi-facet analysis, correlation/association analysis, assessment, comparison, and many other tasks. Secondly, the sophisticated financial data structure increases the difficulty of analysis. The financial systems are usually modeled as temporal financial networks composed of high-dimensional attributed entities and their complex links. This requires a flexible, intuitive, and interpretable representation of sophisticated data. To assist the advanced risk analysis, we need not only the topology but more a comprehensive layout that can support multi-risk analysis such as risk pattern discovery, risk correlation analysis through aggregation, and more importantly, assist prescriptive decisions such as proactive interventions for risk mitigation.

\begin{figure}[!tbh]
  \centering
  \includegraphics[width=0.8\linewidth]{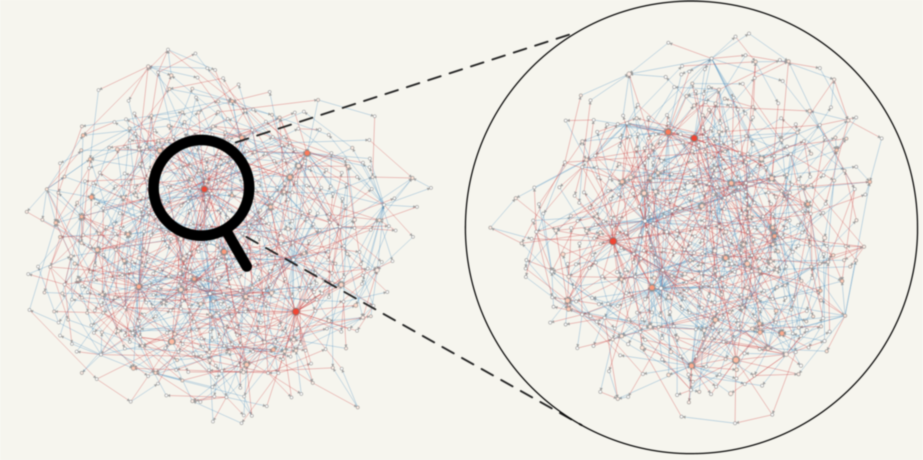}\vspace{-8pt}
  \caption{The 500 node interbank network visualization using force directed graph layout. As can be seen, the nodes are densely connected together, making it difficult to follow the visual clues. For example, the nodes on the left connected to other 482 densely connected nodes (right). In contrast, our risk-island can organize the nodes better (see Figure~\ref{morelarge}a). Although both visualizations have the same visual encoding, the previous popular force-directed graph visualization displays the network with serious visual clutters; users have difficulty discerning the risk patterns, thus, could not practically support risk mitigation Intervention.}\label{fdg}\vspace{-10pt}
\end{figure}

\begin{figure}[!tbh]
  \centering
  \includegraphics[width=1\linewidth]{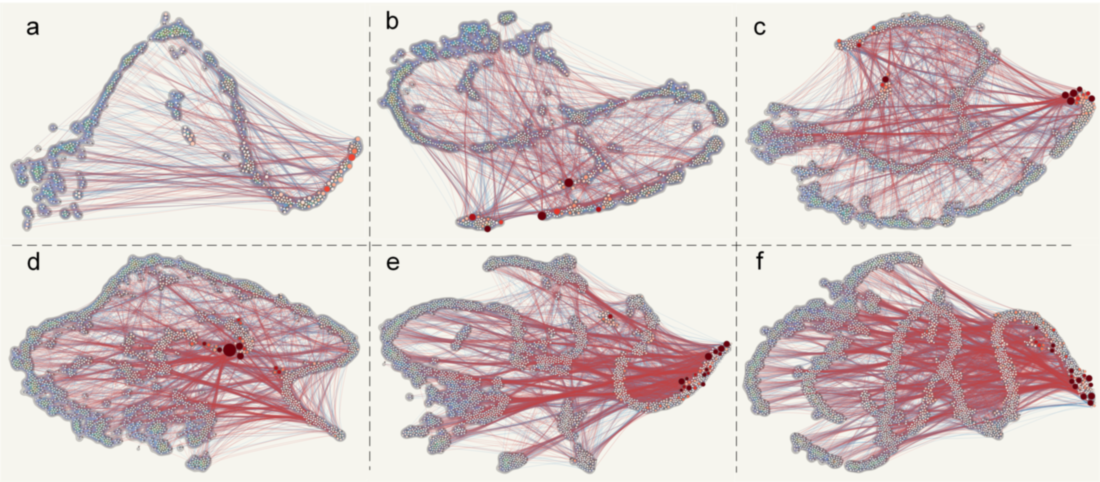}\vspace{-8pt}
  \caption{Risk-island visualization of interbank networks from 500 nodes to 3000 nodes. Zoom in $\geq 2X$ times or see appendix for better view. }
  \label{morelarge}\vspace{-10pt}
\end{figure}

Traditional graph visualizations such as force-directed graph layout are not sufficient to assist these complex business analysis. Figure~\ref{fdg} gives a financial network visualization result using the standard force-directed graph visualization (the node and edge are encoded the same with us in Figure~\ref{morelarge}a). Our financial experts comment that their community widely accepts such visualizations. However, they found that it could deal well with simple networks but could not effectively gain more complex insights and ``got lost'' on large financial networks. These algorithms are designed independently of the business, limiting their ability to support complex problem analysis. And this requires us to think more about ``\emph{business-centered}'' visualization design to support these advanced analysis.
\vspace{-10pt}

The ``\emph{business-centered}'' visualization design requires a confluence effort of business models and data visualizations. In this work, we focus on how to the seamless integrate the financial network visualization with their multi-risk analysis tasks. Work closely together with our financial experts, we identify the  ``\emph{risk-centered}'' financial network visualization design requirements as follows ([BV]: business visualization requirement; [DV]: data visualization requrement): 1) The vertices with similar business-semantics[BR] should be drawn near others[DR]; 2) The topology should be preserved [DR] as they usually a strong business semantics[BR]; 3) The vertices should be distributed to avoid overlapping[DR]; 4) The edge crossings should be minimized[DR]; 5) Pre-defined visual encoding could be supported to reduce the analyst's workload[BR].

Addressing these requirements, we devise \texttt{risk-island} layout algorithm, aiming to provide a comprehensive layout of risk model and network topology. The layout algorithm includes two parts: node placement algorithm, edge layout. With the layout, we also brief describe the visual encoding of the \texttt{risk-island}.

Driven the \emph{risk-centered} financial network visualization design requirements, the nodes of similar risk patterns should be aggregated near each other but not overlap, we formulate the following objective function Eq~\ref{KL}:


\begin{flalign}
\label{KL}
 &C= argmin KL(P\parallel Q)=\sum_{i\neq j}p_{i,j}log\tfrac{p_{ij}}{q_ij}\\
 &p_{ij}=\frac{p_{j|i}+p_{i|j}}{2N} \\
 &p_{j|i}=\frac{exp(-d(x_i,x_j)^2)/2\sigma_i^2}{\sum_{k\neq i}exp(-d(x_i,x_k)^2)/2\sigma_i^2}\\
 &q_{ij}=\frac{(1+\parallel y_i-y_j\parallel ^2)^{-1}}{\sum_{k\neq l}((1+\parallel y_k-y_l\parallel ^2)^{-1})}
\end{flalign}

Where $P$ are the distribution of nodes placements and $Q$ are distribution of their risk models(see Section~\ref{features}). We define the locations of the financial node $y_{i}$ are determined by minimizing the distance (preserving the risk semantics). In this paper, we use Kullback-Leibler divergence to measure the similarities. $p_{ij}$ is the pairwise similarity between nodes $x_i$ and $x_j$ measured using a joint probability. $q_{ij}$ is the similarity between their multi-risks $y_i$ and $y_j$; we employee a normalized heavy-tailed kernel to measure the similarities. It can be minimized by descending along the gradient:\vspace{-8pt}

\begin{flalign}\label{optimizer}
&\frac{\partial C}{\partial y_{i}}=4 \sum_{j\neq i}(p_{ij}-q_{ij}q_{ij}Z(y_i-y_j)) \\
&where\ Z=\sum_{k\neq l}(1+\parallel y_k-y_l\parallel^2)^{-1}
\end{flalign}

\begin{figure}[!tbh]
  \centering
  \includegraphics[width=1\linewidth]{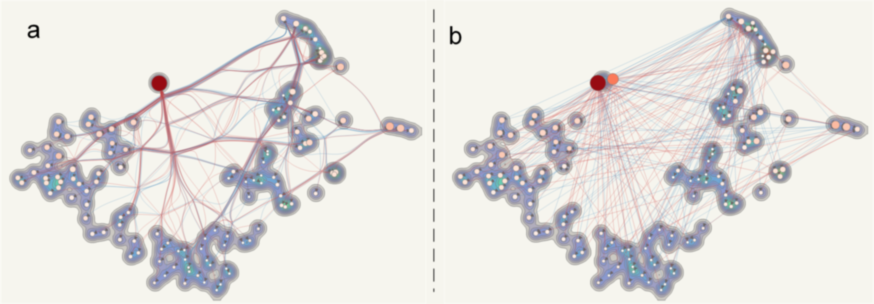}\vspace{-8pt}
  \caption{a, There are coincidences of vertex using Eq1, so we introduce repulsive for the nodes to avoid visual overlaps; b, We use edge bundling technology to assist revealing the pattern of fluidity.}\label{nodeandedge}\vspace{-10pt}
\end{figure}

It is noted that the above algorithm generate nodes placements but might introduce coincidence. For example, as Figure~\ref{nodeandedge}a shows, the two nodes (b55 and b28 refer to Figure~\ref{fig:teaser}) are overlapped. In order to avoid overlapping, we further incorporate the repulsive forces that is defined as $f_r(d)=-k^2/d$, where $d$ is the distance between the two vertices, and $d$ is the radius of the empty area around the vertex. As can be seen from Figure~\ref{fig:teaser} and Figure~\ref{morelarge}, the risk-centered layout can well distribute all the nodes while keeping their risk similarities. We note the clustered area of multiple nodes as ``island'' addressing the map metaphor design decision (D.1). The ``islands'' have different risk-properties, also refer to Section~\ref{case1}.

\textbf{Edge Layout.} We also observed that although we could preserve the topology based on the node placement algorithm, the multiple edges are still challenging for pattern analysis (see Figure~\ref{nodeandedge}b). So, we introduce edge bundling technology to bundle them together assist revealing fluidity patterns.

\textbf{Visual Encoding} In order to fulfill the last visualization requirements, we provide pre-defined visual encodings as well as user configuration windows for the users to define arbitrary encoding meanings to reduce the exploration workload. In this paper, we use the nodes' color to encode to number of defaults, and use nodes' size to encode the strength of stress, and use edges' color to encode liabilities.

We give a serial of examples in Figure~\ref{morelarge} using interbank networks from 500 nodes to 3000 nodes. As can be seen, the risk-island model presents a more organized layout than the traditional ones.

The \texttt{risk-island} layout in-depth integrates the risk model with the graph model to facilitate risk-related analysis. In the \texttt{regshock} approach, the  \texttt{risk-island} overviews the risks at various level and work as a risk-aggregated exploration start point. Users are able to explore multi-risk coordianted with the detail view, conduct intervention operation (through the operating table in Figure~\ref{fig:teaser}b), and assist evaluate the risk mitigation intervention.

\subsubsection{Detail View}
As aforementioned, risk mitigation requires multi-risk assessment and extensive exploration. A practical financial network is represented as a large adjacency matrix. The \texttt{risk-island} visualization provides an ``aggregated'' map of the financial entities according to their risk similarities. It supports navigation and exploration analysis. However, the practical analysis, such as discovering the similarities of financial entities, correlations between risk indicators, requires conduct analysis on the original high-dimensional unaggregated adjacency matrix.

Addressing the requirements, we design the detail view (Figure~\ref{fig:teaser}c), to assist users in exploring more risk details, discover risk patterns, correlations, and clusterings. We choose to use a pixel-based matrix visualization to represent the adjacency matrix. Each row represents a financial entity, and each column represents a normalized risk-indicator. We use the blue-red colors to encode the value of risk-indicators. Reordering according to the column value interaction is supported to view the correlations.

\begin{figure}[!tbh]
  \centering
  \includegraphics[width=1\linewidth]{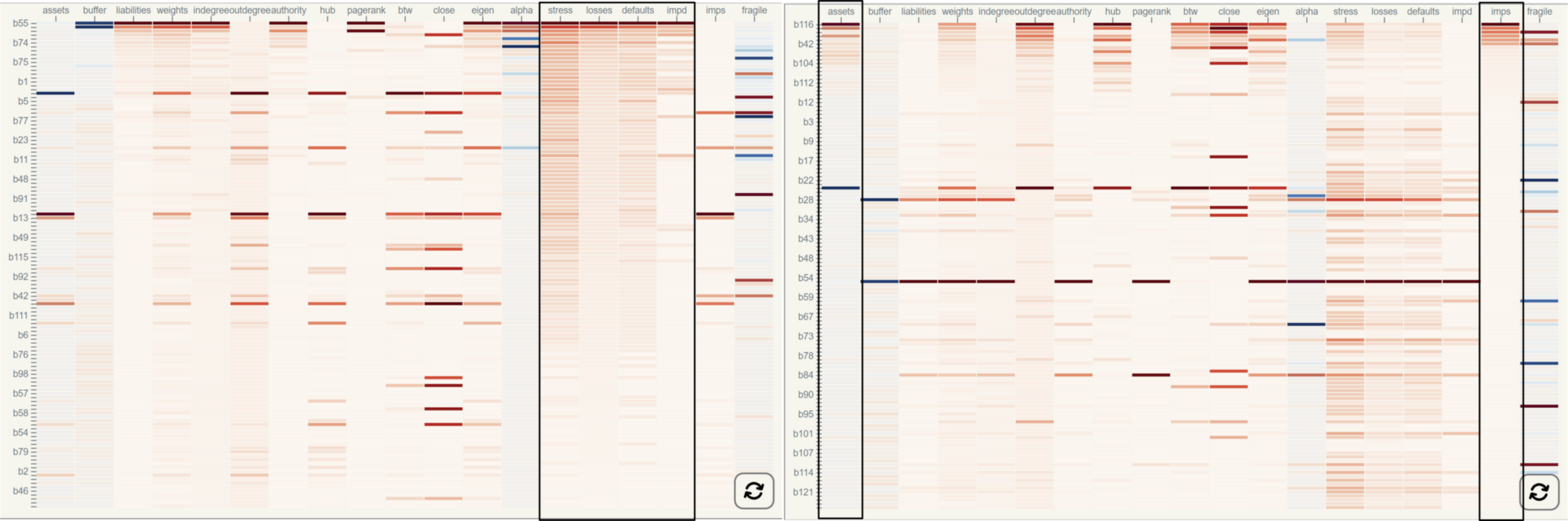}\vspace{-8pt}
  \caption{Detail view for risk pattern analysis and correlation analysis in the intervention strategy S0 in Case2. Left: ranking according the value of stress for correlation analysis of risk indicators; ranking according the value of impact susceptibility for correlation analysis of risk indicators. }
  \label{reorder}\vspace{-8pt}
\end{figure}

Figure~\ref{reorder} gives the reordering results using stress and impact susceptibility. We can observe a strong positive correlation between stress with losses, defaults, impact diffusion, and others; and a strong positive correlation between impact susceptibility and assets. These analysis can enhance the understanding of risks. Besides, coordinated interactions are supported, for example, when users select a group of financial entities in the \texttt{risk-island} view, the selected nodes will be aggregated together on the top of the detail view to facilitate better exploration.

\subsubsection{Intervention Assessment View}
There could be multiple intervention strategies; each may have various consequences to the financial system. To fulfill the simulation-intervention-evaluation analysis loop, we have to provide intervention assessment to help users choose the optimal risk mitigation strategy.

The intervention operations will impact the topology and multi-risks of each entity in the financial networks. Aforementioned, reducing the systemic risk and avoid massive loss is the ultimate goal. The users may consider three aspects for the assessment: 1) intervention cost (rescue cost); 2) system-level risks; 3) entity level multi-risks and their distributions.

\begin{figure}[!tbh]
  \centering
  \includegraphics[width=1\linewidth]{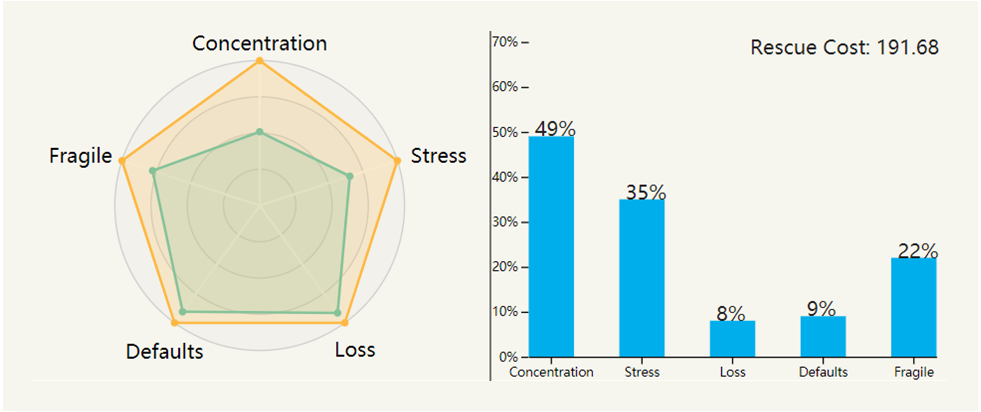}\vspace{-8pt}
  \caption{The intervention assessments are conducted on two alternative views. Left: we use overlapped radar chart to illustrate the systemic risk before and after intervention; Right: we use bar map to character the risk relief percentage and use numbers to illustrate the intervention cost. The result in the above is the Intervention S0 in the Case2.}\vspace{-10pt}
  \label{AssessmentView}
\end{figure}

Addressing the above business requirements and data characterize, we design the intervention assessment view as Figure~\ref{AssessmentView}. It is mainly composed of the systemic risk subview and parallel coordinates entity level multi-risk subview (refer to Figure~\ref{fig:teaser}d and Figure~\ref{AssessmentView}). In the systemic risk subview, five system-level risks (concentration, fragile, maximum stress, total defaults, and total loss) are visualized using a radar chart. We overlay the two radar charts before and after the intervention to assist users in comparing the consequence (Figure~\ref{AssessmentView}a). We also provide an alternative quantified view using a bar map to charting the intervention cost and the risk
relief percentage. In the entity-level multi-risk subview, we choose five risk factors (stress, loss, defaults, fragile, impact diffusion, and impact susceptibility) as the parallel coordinates' axis. The users can observe the distributions, correlations, and values to analyze the entity-level risks. They can also use the switch button (\scalerel*{\includegraphics{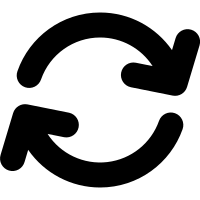}}{\strut}) to exchange the risk data before and after the intervention. These subviews address the financial experts' concerns and help them achieve optimal risk mitigation intervention strategies objectively.

\subsubsection{Interaction Design}
\label{tools}
Interaction tools are supported to investigate risk sources (see left-up corner in Figure~\ref{fig:teaser}a). We have three types of tools:

\setlist{nolistsep}
\begin{itemize} [noitemsep, leftmargin=*]
  \item \textbf{Data operation tools.} Users can load a new interbank network dataset through the file open button (\scalerel*{\includegraphics{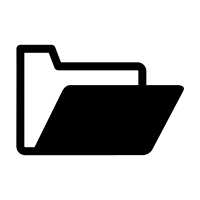}}{\strut}). They may apply various shocks to the interbank network through the shock button (\scalerel*{\includegraphics{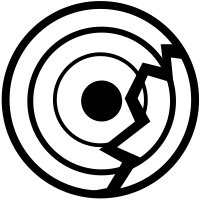}}{\strut}). They can define the shock types (see Section~\ref{datamodel}) and parameters through the configuration panel and apply it to the interbank network through a backend simulation program.

  \item \textbf{Risk exploration tools.} The brush button (\scalerel*{\includegraphics{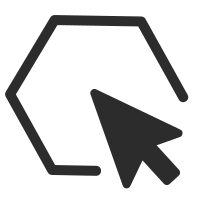}}{\strut}) enables the users to select the nodes/risk-islands of interest and make hypothesis through observing the details (Figure~\ref{fig:teaser}d) and evaluation the shock/partition effect (Figure~\ref{fig:teaser}d). The detail button (\scalerel*{\includegraphics{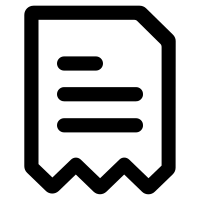}}{\strut}) prompts bank profiles and the focus button (\scalerel*{\includegraphics{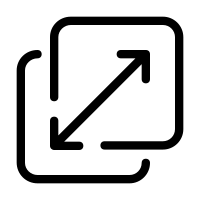}}{\strut}) enables users to view the change of locations on a side-by-side risk-island view. The flow button (\scalerel*{\includegraphics{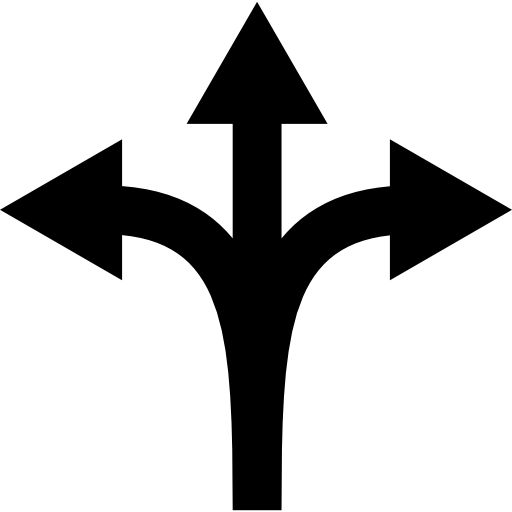}}{\strut}) highlights the liquidity (see anmiation in Figure~\ref{animation}). The color palette button (\scalerel*{\includegraphics{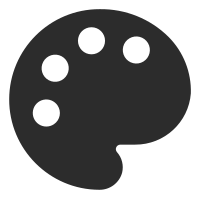}}{\strut}) enables the visual encoding of nodes; we have some preload color patterns to reduce the workload as described in appendix. A detail list button (\scalerel*{\includegraphics{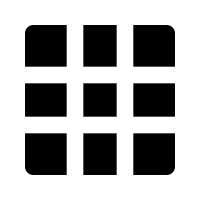}}{\strut}) facility the users to view the original dataset in a normalized heatmap view (refer to Figure~\ref{fig:teaser}). The switch button (\scalerel*{\includegraphics{img/icon/switch.png}}{\strut}) facilite the alternative view of original dataset heatmap visualization, risk-island after partition and details in the heatmap form. The AI button (\scalerel*{\includegraphics{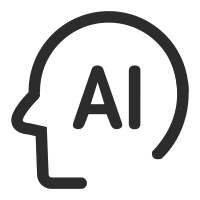}}{\strut}) will also provide hints for risk mitigation operations. A trained graph neural network backend provides risk estimation level are highlighted on the risk-islands (adapted from~\cite{cheng2020contagious}).
  \item \textbf{Risk mitigation tool.} Effectively partitioning the financial network is prominent in mitigating the systemic crisis and building a more resilient financial system to shock. The operation button (\scalerel*{\includegraphics{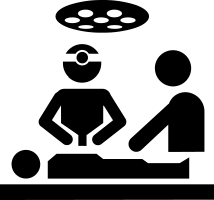}}{\strut}) facilite the users to partition the network on the operating table (Figure~\ref{fig:teaser}b).
\end{itemize}

Besides, coordinated interactions are supported to risk exploration and evaluation. We will show it in the case studies.

%

\section{Evaluation}



We present two use cases to demonstrate how the \text{regshock} system can assist risk mitigation and experts' review to evaluate the approach.

\subsection{Data}
\label{datamodel}


One representative financial network is the interbank market network, which banks use to manage cash demands by shuttling funds among themselves, such as overnight. This process creates a dynamic network of banks linked through the interchange of funds. We utilize the simulated interbank network in our research as common bilateral exposure data are confidential. Notably, the designed visual analytics approach could be smoothly adapted to other kinds of financial networks such as networked-loans~\cite{niu2020iconviz} by aligning the financial behaviors.

The \texttt{regshock} supports two popular interbank network simulations (maximum entropy~\cite{upper2004estimating} and minimum density estimation~\cite{anand2015filling}) to generate the data. The principles to simulation interbanks are based on 1) diversify its exposures~\footnote{Financial term, refer to the amount an investor stands to lose in investment.} as evenly as possible, given the restrictions for both algorithms; 2) sparse and disassortative for the latter one. The simulated interbank network comprises the fictitious bank with the attributes of interbank assets, interbank liabilities, bank capital buffer, and bank weights. In the case studies, we estimate the interbank network through the minimum density estimation model~\cite{anand2015filling}. The simulated network consists of 125 nodes and 249 edges. We apply linear shock propagation to the interbank network~\cite{bardoscia2015debtrank}. 

\subsection{Case Study 1: Multi-risk Exploration}
\label{case1}
In the first case study, we illustrate how to use \texttt{regshock} to explore various aspect risk and provides clues to the risk mitigation operation.

\textbf{Risk-island Overview.} We start to explore the overall risk, as it is usually the start point of analysts. We observe from the parallel coordinates view (Figure~\ref{fig:teaser}) that most of the risk indicators are at the lower cases; however, there are several lines above the axis. Co-considering the detailed heatmap (Figure~\ref{fig:teaser}), we can conclude that the overall risk levels are acceptable, with only a few banks are at a high-risk level. Next, we explore the regional risk patterns using the explore tool (\scalerel*{\includegraphics{img/icon/brush.png}}{\strut}) on the risk-island view. The view d were updated coordinated. We can view from the risk-island map that the banks in the network are divided into multiple risk-islands. The nodes on each island have very similar characteristics in terms of stress, additional defaults, and linkage patterns with other banks. We discovered that the risk-island has at least four risk patterns (Figure~\ref{riskpattern}). They can be summarized as:

\begin{figure}[!tbh]
  \centering
  \includegraphics[width=1\linewidth]{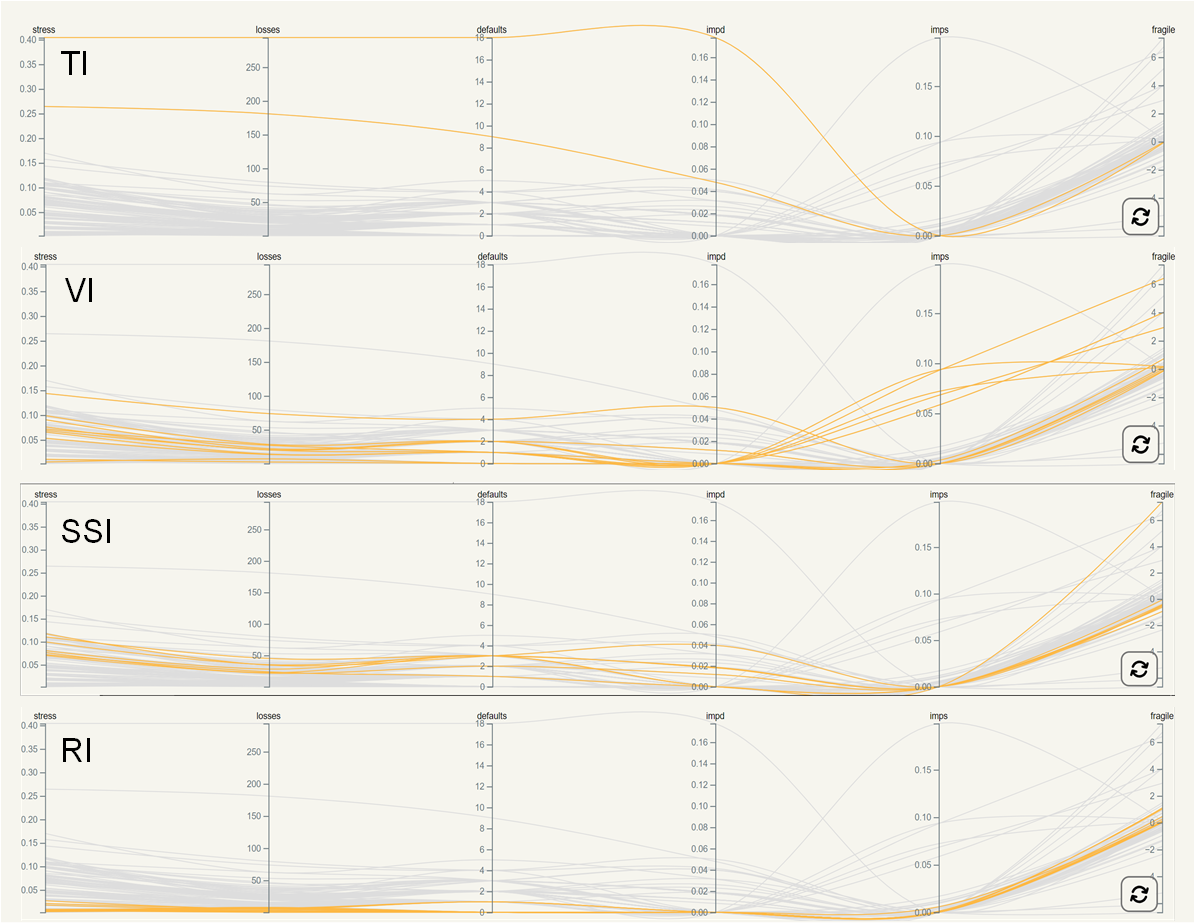}\vspace{-8pt}
  \caption{The risk-island exhibit strong semantics. Based on the multiple risk levels, the islands can be named as: Threatening Island (TI), Vulnerable Island (VI), Suboptimal Status Island (SSI), Resilient Island (RI). Also refer to Figure~\ref{fig:teaser} for detailed explanations.}
  \label{riskpattern}\vspace{-10pt}
\end{figure}

\setlist{nolistsep}
\begin{itemize} [noitemsep, leftmargin=*]
\item \textbf{Threatening Island (TI).} The nodes on TI have high systemic risk values (systemic stress, losses, defaults, and impact diffusion), which means they can pose a substantial threat to the entire financial system. Special attention should be paid to the nodes on this island.
\item \textbf{Vulnerable Island (VI).} The nodes on VI have high vulnerability to other vertices beyond their direct neighbors (remotely vulnerable).
\item \textbf{Suboptimal Status Island (SSI). }The nodes on SSI have risk values moderately; no bankrupt was observed. Although they are affected to some extent by the shock event, the impact is still tolerable. Nevertheless, more attention should be paid to higher-level shocks.
\item \textbf{Resilient Island (RI). }The nodes on RI have a low value of risk aftershock, and this means that they are unscathed and can well digestive the risk brought by the shock. The financial regulators can put them into the ``reassuring'' list.
\end{itemize}

It can be seen that different risk-island exhibit similar risk types. The risk-island visualization can effectively preserve the risk semantics with a relatively organized layout (R.1). It works as a explorer together with the coordinated detail view (Figure~\ref{reorder}) facilitate in-depth analysis.

\begin{figure}[!tbh]
  \centering
  \includegraphics[width=1\linewidth]{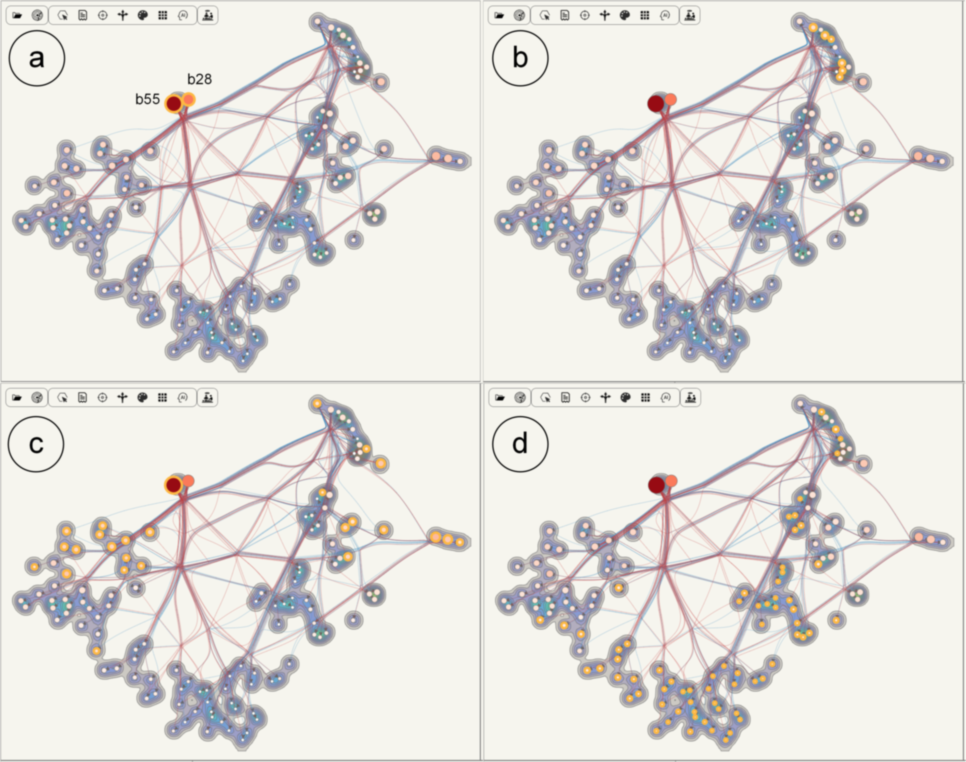}\vspace{-8pt}
  \caption{We are able to gather node lists according to their risk patterns. a, banks with high systemic stress, losses, defaults, and impact diffusion;  b,banks with high susceptibility; c,bankrupt, d, stable banks. It is noticed that there is no coincide between c and d.}
  \label{riskcase}\vspace{-8pt}
\end{figure}

\textbf{Explore and Gathering the Operational Nodes.} We next explore the risk-islands to gather operational nodes that are candidates for future risk mitigation (R.2,R.3). We brush the risk indicator axes of parallel coordinate (Figure~\ref{fig:teaser}d) and highlight the nodes in the risk-islands. This step is similar to the interactions in Figure~\ref{riskpattern}, but in a reverse operation. We can generate several node candidates with different risks as shown in Figure~\ref{riskcase}. Figure~\ref{riskcase}a, we can see that two extremely risky nodes (b55 and b28) are on the same risk-island. By interacting and further observing the detailed information using (\scalerel*{\includegraphics{img/icon/detail.png}}{\strut}), we find that these two banks are extraordinarily indebted and have the highest risk spreading ability in the risk assessment. They are under the most significant risk pressure in this shocking event, which will result in a high capital loss for themselves and also increase the probability of default for connected banks. It is possible that only two banks' failure would cause the entire interbank network system to collapse.

\begin{figure}[!tbh]
    \centering
  \animategraphics[loop,width=1\linewidth]{2}{img/r}{1}{3}\vspace{-8pt}
    \caption{Visual encoding example of risk-islands. Liquidity visualization on risk-islands through mouse over interaction (detail on demand). Click on the image and view the animation using Acrobat PDF reader. A comparison of force directed model with our risk-island model with same color encodings.}
    \label{animation}\vspace{-10pt}
\end{figure}

Besides, we observe that the nodes in Figure~\ref{riskcase}b has high susceptibility, which means the vertex is vulnerable to other vertices beyond its direct neighbors (remotely vulnerable). Figure~\ref{riskcase}c are nodes that have negative fragility values, which means these banks are bankrupt after shock~\cite{battiston2012debtrank}; at last, the nodes in Figure~\ref{riskcase}d are stable ones aftershock as all their risk metrics are in a lower level. In practice, the financial regulators need to intervene the network nodes in Figure~\ref{riskcase}a,b,c and exclude the nodes in Figure~\ref{riskcase}d.

In this case study, by analyzing the systemic level risk and node risk, we show how the risk-centered layout can support risk exploration and operational nodes identification through coordinated views. Compared with traditional force-directed graphs with risk encoding, the risk-island visualization method can intuitively reveal semantic features while preserving the topology (liquidity). The more semantic organized nodes and edges reduce the risk mitigation burden.

\begin{figure}[!tbh]
  \centering
  \includegraphics[width=1\linewidth]{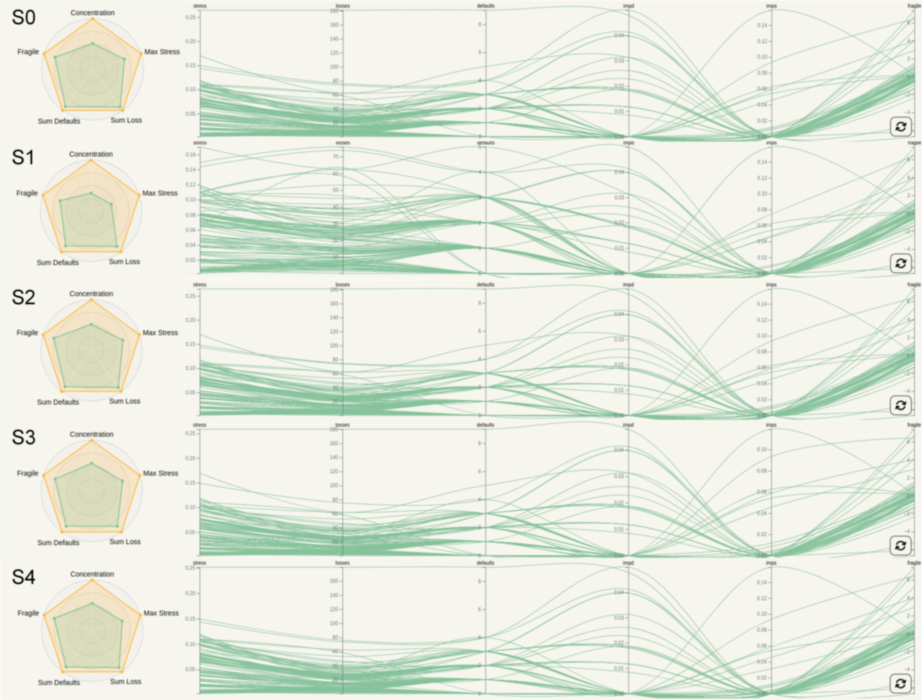}\vspace{-8pt}
  \caption{Strategy evaluation results. Zoom in for the axis details.}
  \label{case.results}\vspace{-8pt}
\end{figure}

\subsection{Case Study 2: Network Partition and Risk Mitigation}
Based on the candidates collected, we are able to conduct network partition and evaluate the effectiveness of risk mitigation.

From case study 1, we observed that the bank (b55) ranks the highest level of systemic risk as shown in Figure~\ref{riskcase}a. However, some other nodes are located in different risk-islands and exhibit various risk patterns (Figure~\ref{riskpattern}). Thus, it is reasonable to remove the node from the interbank network. We use the operation button (\scalerel*{\includegraphics{img/icon/surgery.png}}{\strut}) to cut all the edges in red. This operation means we clear all its debt, which is also the rescue cost. We noted this operation as S0. Since this will impact both the node features and interbank topology, we apply the same configured shock again to generate the new interbank network. The updated risk-island results are given in the visual assessment view (right-up corner) in Figure~\ref{method}.

We can observe that the new risk-island consists of more loosely distributed nodes with lower risk concentration. We further observed through offline examinations that 43 nodes slightly increased their stress level after removing b55. Among them, 17 nodes have a direct link to node b55 while others are not. This means operations to any entities in the network will impact the financial system. Then we look for more improvements, so we continue to explore operations to risk mitigation. The candidates generated from the previous explorations provide us more clues and evidence. We further remove nodes in a high-risk level besides b55. We summarize the following four strategies:

\setlist{nolistsep}
\begin{itemize} [noitemsep, leftmargin=*]
\item \textbf{S1: remove high systemic risk nodes.} We further remove b28, which is locates same with b55 on TI.
\item \textbf{S2: remove moderately risk nodes.} we further remove b5, which locates close to b55 but on SSI.
\item \textbf{S3: remove high susceptibility nodes.} we further remove b33 (on VI). b33 has the third-highest number of defaults and level of stress except for b55 and b28; it also has a high level of susceptibility which means it vulnerable to other nodes (refer to Figure~\ref{riskpattern}b and Figure~\ref{riskcase}b). Node b28 does not directly link to bank b55.
\item \textbf{S4: remove bankrupt nodes.} we further remove b74. b74 has a large number of defaults and a level of stress, and most importantly, it is one of the bankrupt ones aftershocks (refer to Figure~\ref{riskcase}c).
\end{itemize}

\begin{figure}[!tbh]
  \centering
  \includegraphics[width=1\linewidth]{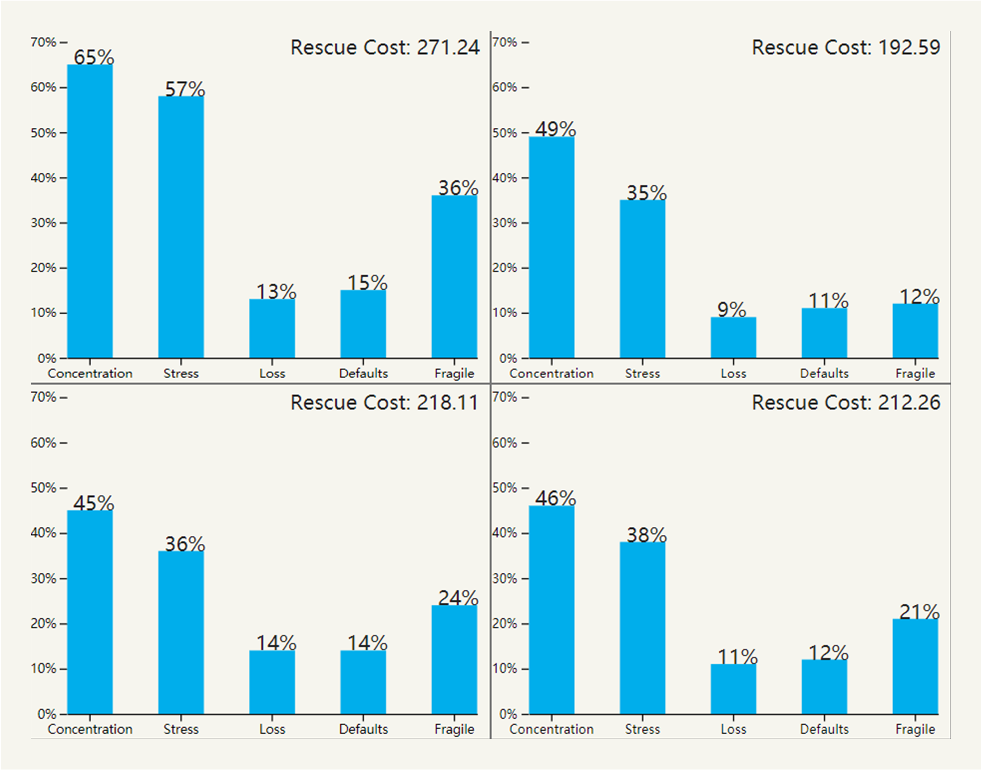}\vspace{-8pt}
  \caption{Risk relief of different strategies through surgical network partition operation. Rescue cost in unit, reduction in percent.}
  \label{relief}\vspace{-8pt}
\end{figure}

We give the risk mitigation results in Figure~\ref{case.results}. The rescue cost and risk relief are given in the intervention assessment view as shown in Figure~\ref{relief}. We can see that all these interbank intervention operations can relieve systemic risk to a certain extent (R.3, R.4). Among these strategies, removing the highest systemic risk node has huge resilience improvement of the financial systems. For example, in S0, removing merely b55 can reduce almost half of the systemic concentration, one-third of maximum stress, and one-fifth systemic fragile level. However, further removing the highest systemic risk nodes will face a more rescue cost. For example, in S1, removing node b28 means the financial system needs to take an extra 41\% cost with a moderate risk relief. Removing moderate-risk nodes does not help (S2); removing high susceptibility nodes (S3) can save more systemic loss but with extra cost, and it has a similar effect to the operation of removing bankrupt nodes (S4).

We demonstrate the strategy exploration and assessment capability of risk mitigation through the case study. It is noted that we only list several operational strategies with results here. The financial regulators are responsible for judging from this evidence and making proper decisions on choosing the best operations.

\subsection{Expert Review}
We conducted expert review to further \texttt{regshock} evaluate how well users could understand the metaphors and exploit it to mitigate risk.

We invited 8 financial experts. After a 30 minutes online tutorial, they were asked to explore the multi-risk and try to mitigate risk through our online demo. Questions are encouraged. After that, we conducted a semi-structured interview guided by the questions below. Overall the feedback is positive and encouraging, we also received some great suggestions from them. We have adjusted our approach according to their comments. Our interview guidance are: 1) Significance. is the problem is critical? How do you think about the interactive methodology for complex financial problems? 2) Effectiveness. Is it easy to explore and understand the risks in the financial network (sense-making)? Is it easy to verify your risk mitigation hypothesis through the system (decision-making)? 3) Visual and Interaction Design. Is it easy to understand the metaphors in the risk-island? are the interactions natural and in line with analytical thinking habits? Do you have any other suggestions?
We summarize the feedback and our response as follows:

\textbf{Significance:} The importance got affirmed by all interviewees. Systemic risks have always been highly concerned by central banks and regulatory officials.  They were all glad to see such an intuitive and interactive approach to examine their ideas. One expert commented that ``we saw a trend of interdisciplinary effort understanding risk through human-machine co-operation in the financial regulation business; such a visual analytics approach potentially provides us with actionable insights.''. Another financial expert (knowledgeable in the research of networked-loan problem) complement that, although our case studies are conducted on interbank network data, it seems the approach could be applied to the networked-loan problem, the one long plagued financial regulatory authorities (also refer to our work~\cite{niu2020iconviz, niu2018visual}). They used to follow a principle of differential measures for preventing risk but lack practical instructions. They have to purely rely experts' experience that sometimes has unpredictable results. The \texttt{regshock} can be adapted to the networked-loan problems and facilitate them a dispose handle. We agree with these suggestions and will extend our \texttt{regshock} approach to the more complex financial regulatory problem in our future work.

\textbf{Effectiveness:} The simulation-intervention-evaluation analytic methodology for hypothesis validation attracted their great attention. This work roots from some real systemic risk management concerns, which are highly related to the experts' domain. They all remarked that it was beneficial to visualize the risk clues and were delighted that they are able to verify their ideas and get feedback intuitively. Some of them attempted to mitigate the risk in the financial network (networked-loans); however, their operations are based on experience by asking some high seniors to replace critical nodes and take responsibility in the network (by introducing more bank financial support). There could be some more ``low-cost'' possibilities,  but they have difficulty assessing the strategies. On the other hand, they could usually only analyze relatively small networks. Thus, it was not easy to immerse themselves in the massive body of data and analyze their interest topics. The regshock system granted them the ability to avoid this dilemma. They did not get lost in an ocean of network data. They could quickly locate their risk of interest and performed in-depth analyses.

We identified several usability issues through the expert interviews. For example, in our initial version, the system provides only risk-island visualization of the entire financial network; sometimes, they would like to access the original data attributes. We agree with the suggestions and have adjusted our system by adding the detailed heatmap view (Figure~\ref{fig:teaser}b). The users are able to alter different views (detailed heatmap view of original network and the one after surgical partition, and the risk-island view after partition). A second suggestion is to use qualitative results like the charts and hope to see some quantitative analysis in a practical scenario. In response to this issue, we have added another bar map to give the results of quantified risk mitigation effects (see Figure~\ref{relief}). Besides, one risk control manager said that deep learning-based risk prediction is popular in bank risk management systems these days. He suggested that the current \texttt{regshock} could consider some more intelligence assistance. We have achieved some state-of-the-art deep graph neural network-based approach~\cite{cheng2020contagious}, and we will keep incorporating such artificial intelligence advance in to our visual analytics works. We were glad to receive this constructive suggestion and plan to improve the \texttt{regshock} approach.

\textbf{Visual and Interaction Design:} The experts expressed very positive comments of the overall design. The light-yellow background inspired a sense of spirituality and encouraged creativity. What they most appreciated was the risk-island idea. A practical financial network could consist of hundreds or even tens of thousands of nodes in the extreme case, which brought them a significant burden. They used to access force-directed graph visualization, and it is pretty difficult to gain insights about the risk level as there are so many factors, so they used to use it only fore information communication but not business analysis. With such an intuitive layout, they did not need to use the mouse wheel to examine nodes one by one and get lost in the data (their words). They liked the risk-semantic organized layout, compact, and informative design of the views. One expert commented that the 125-node financial network in the case study is a medium complexity one. The system's interactions are very smooth, and they would like to know whether our system could deal with more complex financial networks. Other experts go along with these comments. After the interview, we take the experiments and provide more results to them. Generally, our system has a good scaling performance on risk-semantic visualization and interaction Intervention operations. We also provide the risk-islands of large financial networks (500-nodes, 1000-nodes, 1500-nodes, and 2000-nodes). There is a slight interaction delay for large networks; however, we believe this is because of the SVG set. Our approach could be smoothly transferred to WebGL-based solutions, which will have no difficulty handling networks over 10,000 nodes.

The experts believed that the interactions in the views presented by our system were beneficial and could help them explore and analyze issues important to their work. The interactions among multiple coordinated views facilitated the closed-loop analysis process and iterative level of detailed exploration. Moreover, there were several functional buttons and selection/zooming interactions supported in each view. One expert said that these were powerful but made the system a bit complex. However, he agreed that it was difficult to make a proper trade-off between complex functions and powerful analytical abilities. For example, he would have had no idea how to use the coordinated views to perform the analysis loop without the walkthrough training, so he suggested that training would be necessary for any future study or deployment. Future work will keep improving the interaction performance.

\section{Conclusion}
In this research, we present our progress of visual analytics of the risk mitigation problem. Addressing the complexity of business analysis tasks and financial network data characters, we propose a novel risk model driven ``risk-island'' visualization and develop a simulation-intervention-evaluation approach (\texttt{regshock}) to facilitate practical multi-risk exploration and assist preventive intervention decision makings. We believe this is the first visual analytics approach for financial risk mitigation problem. We extensively evaluated the approach through case studies and experts' review. We have adjusted our approach based on their feedback. In the future, we will keep improving the ``risk-island'', for example to further reduce the crossings between edges and ``islands'' to reduce overlaps. The current system is evaluated on a serial of simulated interbank network data. We will extend it to other general financial networks such as the urgent networked-loan risk mitigation problem and use real-world data to evaluate the approach.
\bibliographystyle{abbrv-doi-hyperref-narrow}

\bibliography{ref}
\end{document}